%% file: main.tex
\documentclass[manuscript]{acmart} 

\AtBeginDocument{%
  }

\usepackage{pifont}
\usepackage{listings}
\usepackage{soul}
\begin{document}

\title[Limitations of the LLM-as-a-Judge Approach for LLM Evaluation]{Limitations of the LLM-as-a-Judge Approach for Evaluating LLM Outputs in Expert Knowledge Tasks}



\author{Annalisa Szymanski\textsuperscript{*}}
\email{aszyman2@nd.edu}
\affiliation{
\institution{University of Notre Dame}
\city{Notre Dame, IN}
\country{USA}}

\author{Noah Ziems\textsuperscript{*}}
\email{nziems2@nd.edu}
\affiliation{
\institution{University of Notre Dame}
\city{Notre Dame, IN}
\country{USA}}

\author{Heather A. Eicher-Miller}
\email{heicherm@purdue.edu}
\affiliation{
\institution{Purdue University}
\city{West Lafayette, IN}
\country{USA}}

\author{Toby Jia-Jun Li}
\email{toby.j.li@nd.edu}
\affiliation{
\institution{University of Notre Dame}
\city{Notre Dame, IN}
\country{USA}}

\author{Meng Jiang}
\email{mjiang2@nd.edu}
\affiliation{
\institution{University of Notre Dame}
\city{Notre Dame, IN}
\country{USA}}

\author{Ronald A. Metoyer}
\email{rmetoyer@nd.edu}
\affiliation{
\institution{University of Notre Dame}
\city{Notre Dame, IN}
\country{USA}}

\thanks{\textsuperscript{*}Equal Contribution.}

\begin{abstract}
The potential of using Large Language Models (LLMs) themselves to evaluate LLM outputs offers a promising method for assessing model performance across various contexts. Previous research indicates that LLM-as-a-judge exhibits a strong correlation with human judges in the context of general instruction following. However, for instructions that require specialized knowledge, the validity of using LLMs as judges remains uncertain. 
In our study, we applied a mixed-methods approach, conducting pairwise comparisons in which both subject matter experts (SMEs) and LLMs evaluated outputs from domain-specific tasks.  
We focused on two distinct fields: dietetics, with registered dietitian experts, and mental health, with clinical psychologist experts. 
Our results showed that SMEs agreed with LLM judges 68\% of the time in the dietetics domain and 64\% in mental health when evaluating overall preference. Additionally, the results indicated variations in SME-LLM agreement across domain-specific aspect questions. 
Our findings emphasize the importance of keeping human experts in the evaluation process, as LLMs alone may not provide the depth of understanding required for complex, knowledge specific tasks.  We also explore the implications of LLM evaluations across different domains and discuss how these insights can inform the design of evaluation workflows that ensure better alignment between human experts and LLMs in interactive systems.
\end{abstract}




\maketitle

\input{1-introduction}
\input{2-related_work}
\input{3-method}
\input{4-results_SME}
\input{5_lay_user_followup}
\input{6_discussion}

\input{7-conclusion}

\input{8-acknowledge}

\bibliographystyle{ACM-Reference-Format}
\bibliography{bibliography}

\pagebreak
\input{9-appendix}

\end{document}

%% file: 1-introduction.tex
\section{Introduction}

The ability of large language models (LLMs) to follow natural language instructions allows for an enormous number of downstream use cases in the real world~\cite{bommasani2021opportunities, zhao2023survey}. Recently, there has been significant research interest in leveraging LLMs for complex knowledge-based tasks in professional fields such as healthcare, mental health, legal issues, and more~\cite{rasal2024navigating, ge2024openagi}. These tasks often involve question answering and task generation, providing critical benefits, particularly in areas where access to the expertise of trained professionals is limited or where underprivileged communities may lack necessary services~\cite{kirk2021precision}.


However, it remains to be seen how to best evaluate the quality of model outputs for these tasks.
Traditional automated metrics such as BLEU, ROUGE, and Exact Match are commonly used to assess short, well-defined text generations consisting of only a few tokens~\cite{papineni2002bleu, lin2004rouge}.
These metrics become less useful for open-ended tasks where multiple valid responses exist.
Other approaches, such as BERTScore, measure semantic overlap in meaning between texts but fail to capture user preferences, domain-specific nuances, or the complexity of reasoning required for tasks that need expert judgement~\cite{zhang2020bertscore}.
As a result, many tasks requiring open-ended generation rely on human evaluators to judge the correctness of a language model's output. While effective, human evaluation is often expensive, slow, and inconsistent in terms of preferences~\cite{zheng2024judging}.


To address these challenges, the use of LLMs as evaluators has gained popularity within the NLP and HCI community, as they provide an inexpensive and reproducible evaluation compared to human evaluators \cite{ouyang2022training}. These approaches typically involve prompting the LLM evaluator, often referred to as a judge, to perform a pairwise comparison between the outputs of different language models or a single-point evaluation using specific criteria~\cite{arawjo2023chainforge, desmond2024evalullm}.
Previous studies have shown that LLMs are a promising alternative to human expert evaluations \cite{taori2023stanford, zheng2024judging, dubois2024alpacafarm}.
However, there have been several problems with using LLMs as evaluators, such as positional bias, knowledge bias, and format bias~\cite{zhu2023judgelm}.
Although there are approaches to solve these issues, such as randomizing the position of the responses, it remains to be seen whether they can be completely eliminated~\cite{zhu2023judgelm}.
This challenge is further complicated by tasks that require complex reasoning or cognitive processes. Evaluating such tasks often involves a combination of cognitive strategies including analytical reasoning, intuitive thinking, and pattern recognition, while incorporating elements of practical expertise \cite{vo2021role}. This complexity adds a layer of difficulty in both the performance and evaluation of LLM outputs for these tasks. 

This presents a critical gap in understanding how to best assess the performance of LLMs for domain-specific complex tasks, as it remains unclear when and how to rely on SME evaluations versus LLM-based assessments, and how to effectively integrate the strengths of both.
Misalignments between LLM outputs and expert judgments can lead to missed opportunities to integrate LLMs into human decision-making processes, which could otherwise reduce costs and enable faster iterations. The lack of clear guidelines on how or when to integrate LLMs or SMEs into the evaluation process could lead to both resource misallocation or sub-optimal performance in high stake domains, such as healthcare. Additionally, understanding the boundaries of LLM capabilities and their alignment with expert standards is essential for developing more reliable evaluation pipelines that can balance cost effectiveness with the precision required for expert-level tasks.






In this paper, we aim to understand the alignment between LLMs and SMEs as judges for domain-specific tasks that require complex reasoning and cognitive processes.
Our work is driven by the following research questions:
\begin{itemize}
\item RQ1: How does the LLM-as-a-Judge evaluation approach compare to the evaluations conducted by SMEs for domain specific tasks?
\item RQ2: What are the main factors contributing to the evaluation differences and associated explanations between LLMs and SMEs?
\end{itemize}

To answer our research questions, we conduct a case study in the fields of mental health and dietetics where the tasks are complex and may require expert knowledge.
These fields are of particular importance because they represent complex domains that could bring significant societal benefits by improving access to their services. However, a lack of understanding of how LLMs perform on these tasks presents a barrier to adoption. 
Through a mixed-methods approach, we conducted a pairwise comparison study where both SMEs and LLMs evaluate outputs related to domain-specific tasks and provide explanations for their choices. 

Based on our analysis, our findings contribute new insights into how the LLM-as-a-Judge evaluation approach compares to evaluations conducted by SMEs and the variability seen between different expert-level tasks. We observe patterns in LLM agreement and discuss potential reasons for them, as well as key design implications of LLM evaluations for domain-specific tasks.  

This paper makes the following research contributions: 

\begin{itemize}
\item We contribute empirical evidence highlighting the critical role of incorporating SMEs into evaluation processes for complex, domain-specific tasks.
\item We provide analysis of explanations provided by both LLMs and SMEs and offer insights into how and why their evaluations differ across tasks.
\item We discuss the variability in agreement between LLMs and SMEs, emphasizing how this misalignment varies by domain and task type.
\item We design implications for effectively integrating SMEs into the evaluation process, identifying key areas where expert input is essential for improving alignment and ensuring accurate assessments.
\end{itemize}

%


%% file: 2-related_work.tex
\section{Background}

\subsection{LLM Evaluators/LLM as a Judge:}

How to properly evaluate language model outputs remains an open area of research \cite{chang2024survey}.
For tasks such as multiple choice question answering where only a few choices are valid, exact match (EM) or likelihood are often used \cite{talmor2019commonsenseqa}.
However, evaluating generations with more than a few tokens presents new challenges as there are often a large number of valid and correct answers.
For shorter generations such as question answering, token overlap between the generated answer and the labeled answers are measured with F1, BLEU, ROUGE, and others \cite{papineni2002bleu, lin2004rouge}.
These token overlap metrics often struggle when a generated answer is correct but not in the list of labeled answers, as in the case with paraphrasing.
Semantic evaluators such as BERTScore seek to remedy this by measuring the semantic similarity between the target text and the generated text \cite{zhang2020bertscore}.

However, for tasks such as instruction following where a language model generates many paragraphs of text, evaluation is even more difficult as there is an extremely large set of possibly correct answers.
To make matters worse, some sentences within a single generation may be correct while others are not.
For these tasks, a labeled ``correct'' answer is often foregone entirely and instead replaced with human pairwise evaluation, where a human is presented with two different language model generations for the same instruction and tasked with picking which generation better adheres to the instruction \cite{chiang2024chatbot}.
Language models are then compared via a relative metric such as ELO or by win-rate\cite{elo1978rating}.
However, this approach is expensive, time consuming, and difficult to reproduce.
Recently, many works have proposed instead using a separate language model, often called a judge or Evaluator, to address these issues by replacing the human evaluator \cite{zheng2024judging, li2023alpacaeval, dubois2024alpacafarm}.
For example, EvalLM allows users to compare prompts while the LLM evaluates the pairs using predefined criteria, offering transparency by displaying both the number of times a prompt won and the reasons behind the evaluation \cite{kim2023evallm}. EvalGen goes further by enabling users to define evaluation criteria and grade outputs, aligning LLM evaluations with human-defined benchmarks \cite{shankar2024validates}.
Approaches that use LLMs as evaluators, rather than humans, demonstrate a surprisingly high correlation with human preferences \cite{dubois2024length}. Despite these advances, it remains uncertain whether LLMs can align with SMEs in specialized or complex domains.

\subsection{Application of Language Models in Expert Domains}
Language models have demonstrated a promising performance in several domains requiring expert knowledge.
In the mental health domain, ChatCounselor trains a language model on conversations between clients and professional psychologists, achieving performance comparable to GPT-4~\cite{liu2023chatcounselor}. Similarly, Mental-LLM evaluates various open-source and closed-source models for mental health use cases, showing that instruction tuning a smaller model can outperform larger, general-purpose models~\cite{xu2024mental}.

New approaches are also emerging in the dietetics domain. ChatDiet introduces a framework for using LLMs to provide personalized, nutrition-focused food recommendations~\cite{yang2024chatdiet}. Likewise, \citet{szymanski2024integrating} offer guidelines for developers on prompting LLMs to deliver high-quality nutrition information. Additionally, \citet{niszczota2023credibility} investigate the credibility of nutrition advice generated by ChatGPT for individuals with food allergies.





%% file: 3-method.tex
\section{Methods}

\subsection{The Need for Subject Matter Expertise}
To investigate expert knowledge in relation to ranking outputs of complex tasks, we selected subject matter experts (SMEs) from the dietetics and mental health domains. The deployment of LLMs in these fields is increasing, expanding the potential for innovative healthcare solutions~\cite{ji2023rethinking, szymanski2024integrating}. In these domains, decision making is based on clinical judgement and evidence-based practices~\cite{vo2021role, michael2019clinical, martin1999influences}. This type of reasoning integrates a decision-making process that involves weighing evidence, employing a variety of cognitive strategies, from analytical reasoning to intuitive pattern recognition, and incorporating elements of intuition and practice wisdom~\cite{vo2021role, michael2019clinical}. 

Due to inherent biases in LLMs and their reliance on patterns in training data, LLMs may struggle to accurately assess expert-level decision making, potentially overlooking critical factors in complex subject matters~\cite{chen2024humans}.
We argue that this level of expertise is crucial when evaluating LLM outputs, as LLMs may lack the ability to fully capture the complexity involved in complex tasks. 
In addition, while lay persons may possess critical thinking skills, they lack the depth of understanding required to make clinical judgements accurately~\cite{vo2021role}. An average LLM user may not recognize inaccuracies or the importance of certain dietary restrictions and mental health protocols and is less likely to be aware of the latest research and clinical guidelines.


\subsection{Dataset Curation}
\label{Dataset_Curation}
We crafted a dataset of 25 instructions for both the dietetics and mental health domains to test the evaluations performed by both the LLM and the SMEs. This dataset was designed to provide a diverse range of instructions and questions that would challenge both the ability of the LLM to provide accurate and complex responses and the clinical judgment of the SME. 

\subsubsection{Dietetics Instructions}
We surveyed the nutrition and dietetics literature that explore the use of LLMs for nutrition advice on common prompts or questions for dietitians that can be answered by LLMs \cite{szymanski2024integrating, kirk2023comparison, ponzo2024chatgpt, garcia2023chatgpt}. From this survey, we compiled common themes for instructions related to disease management and guidance, dietary preferences and lifestyle, nutrition components and overall health, and food allergies and sensitivities. 
The prompts consider the top diseases and allergies formed in the nutrition literature \cite{kirk2023comparison}. 

\subsubsection{Mental Health Instructions}
Building on a review of LLMs in mental health care, we referenced the work of ~\citet{hua2024large} to compile prompt instructions based on common application areas in LLMs such as conversational agents, emotional support, and on-demand online counseling. From this review, we identified frequently addressed mental health topics, including stress, anxiety, depression, suicide, bipolar disorder, PTSD, autism spectrum disorder, and others. Using these insights, we crafted instructions based on common inquiries from the literature, ensuring that the dataset covered a comprehensive range of mental health concerns.

\subsubsection{Aspect Evaluation Questions}

For each instruction, we developed a set of aspect questions aimed at evaluating specific dimensions of the responses provided. These aspect questions were inspired by previous work on evaluation criteria in the domains of dietetics and mental health~\cite{szymanski2024integrating, doherty2010design}. Drawing from the literature, we identified key areas such as accuracy, clarity, educational context, personalization, and professional standards. Based on these themes, we formulated specific aspect questions, which we group as shown in Table \ref{fig:aspect_questions_method}.

\begin{table*}[!t]
\centering
\includegraphics[width=\textwidth]{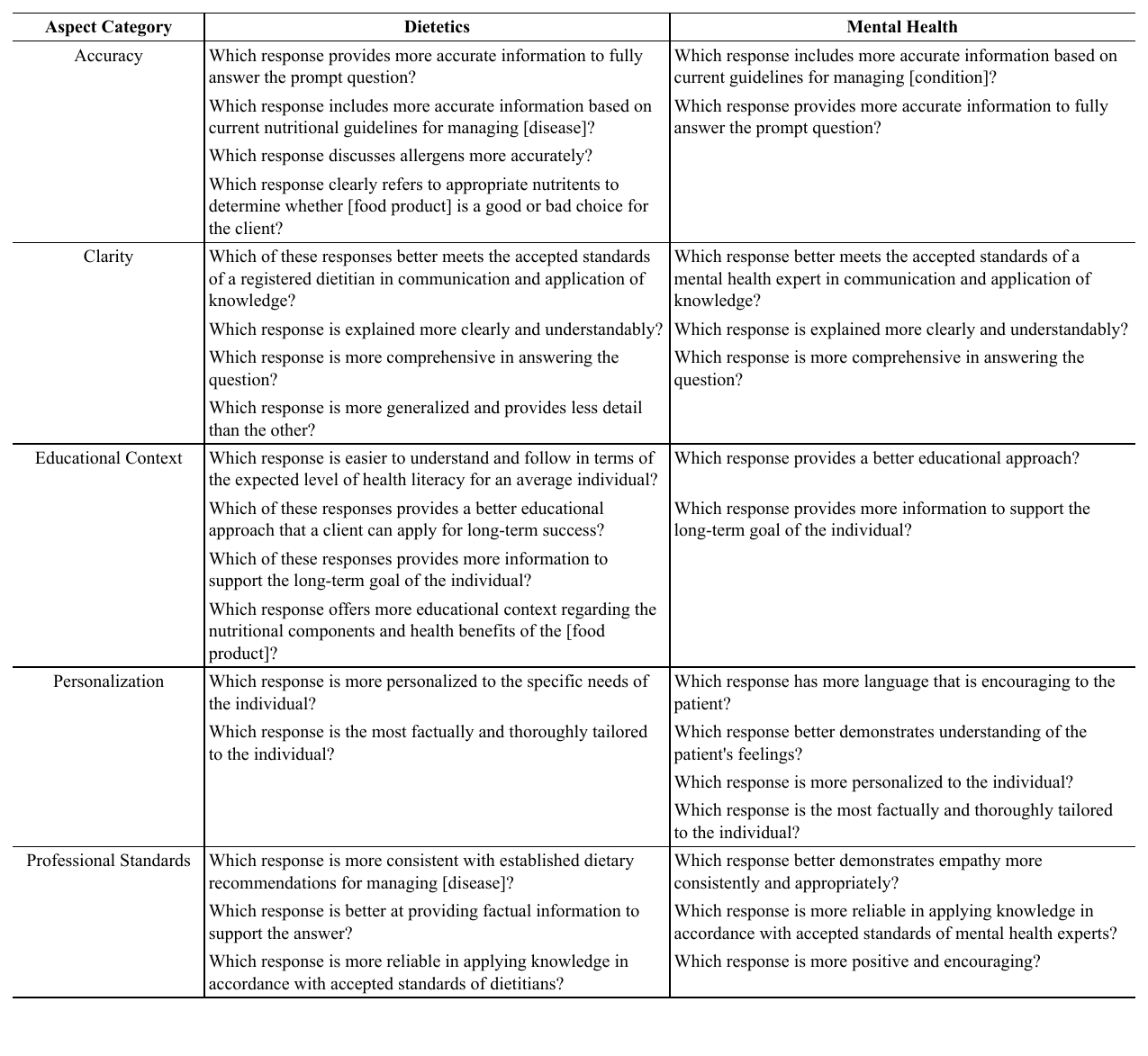}
\caption{Breakdown of Specific Aspect Questions by Category}
\Description{Add description here}  
\label{fig:aspect_questions_method}
\end{table*}

\subsection{Pairwise Comparison Method for LLM Output Quality Evaluation}
We adopt a pairwise comparison method in which both the LLM and the SME is presented with two candidate outputs for the same instruction and is tasked with evaluating which model's output is better \cite{zheng2024judging}. This pairwise comparison method is similar to the commonly used state-of-the-art methods in the Natural Language Processing (NLP) community for evaluating and comparing LLM performances, such as Chatbot Arena, MT Bench, AlpacaEval, and others~\cite{lin2024wildbench, li2023alpacaeval, zheng2024judging}.
This method enables direct comparison between LLM and SME judgments on task-specific outputs.

As in MT Bench and AlpacaEval~\cite{li2023alpacaeval}, to determine the preferences of LLM judges, we use a pairwise comparison approach, in which the LLM judge is presented with two candidate outputs and is prompted to indicate which candidate is better.
An example of the prompt used for evaluation can be found in Table \ref{tab:alpacaeval-template} of the Appendix.
The response with the higher overall likelihood (higher log probability) is considered the better one.
This process is done both to measure overall LLM judge preferences as well as to measure preferences with respect to specific aspect questions.

We also experiment with an additional condition where we give the LLM judge an \textit{expert persona} by indicating that it should personify an expert, to determine whether there would be an impact on agreement rates with the SMEs. Previous research has shown that expert personas enhance the perceived quality of the output~\cite{liu2024personaflow}.
This technique is commonly used to generate more personalized and context-aware dialogue, mimicking the knowledge and experiences of the specified expert in the output generation~\cite{li2023chatharuhi, liu2024personaflow, salemi2023lamp}.

To compare the preferences of LLM judges with SMEs, we take the same set of instructions and candidate generations, but instead task \textit{subject matter experts}, in this case registered dietitians or clinical psychologists, with judging which generation candidate is better.

%% file: 4-results_SME.tex
\section{Experiment}
\label{sec:experiment}

\subsection{Subject Matter Expert Recruitment}
Our work was approved by the Institutional Review Board (IRB) of our university prior to recruiting participants. To reflect the specialty of the domains, we recruited ten registered dietitians and ten clinical psychologists to complete the evaluation survey tailored to the dietetics and mental health domains, respectively. These two groups were chosen for their education and experience in their respective fields. Participants were recruited through email from a network of university campuses, hospitals, and local community practices. All participants were compensated with a \$75 gift card at the end of the study. 

Before starting the survey, participants were asked to consent to participate. In addition, we presented participants with a pre-study questionnaire, through which we gathered information about their educational background, specialty, workplace setting, tenure in the profession, and familiarity with AI. The profiles of the dietitian participants and clinical psychologists can be found in Table \ref{tab:combined_profiles}.


The dietitian participants had an average of 9.6 years \textit{(SD = 8.87)} of experience in the field, while the psychologists had an average of 15.2 years \textit{(SD = 9.78)}. Three out of ten dietitians had previously used AI tools in their practice or research, whereas none of the psychologists reported using AI in their work. When asked to rate their familiarity with AI tools on a scale from 1 to 5 (with 5 being the most familiar), the dietitians reported an average score of 2.8 \textit{(SD = 0.92)}, while the psychologists reported an average score of 2.0 \textit{(SD = 0.67)}. 

\input{tables/participant_profiles}

\subsection{Selected Models}
To create the two LLM candidate responses whose quality will be evaluated in the experiment, we use GPT-4o, and GPT-3.5-turbo ~\cite{achiam2023gpt, ouyang2022training}.  
While there are many models to choose from, we find that GPT-4o and GPT-3.5-turbo are commonly cited in the recent literature \cite{cheng2023now, king2023introduction, liu2023chatcounselor}. 
We intentionally keep this set of models to a limit of two models, as adding more significantly increases the cost of human annotation. 
For all responses, we used a temperature of 1.0.

For the LLM used to judge the quality of instruction following (the judge model), we use GPT-4 as it is widely available and is frequently used as an LLM judge \cite{li2023alpacaeval}.
As with SME participants, we prompt the language model judge to provide reasoning for preferring one response over another.
The in-depth analysis of this reasoning and its comparison with SME reasoning can be found in Section \ref{subsec:sme-v-llm-explanations}.

\subsection{Subject Matter Expert Evaluation Experiment Procedure}
To facilitate the preferences of the SME judge, we construct a survey with a total of 25 evaluations from our curated dataset.
For each of the evaluations, we provide the prompt instruction along with two candidate responses from two different language models.
For each evaluation, participants were first tasked with choosing their general preference through the question \textit{``Which response was better overall?''}. In addition, each evaluation included two aspect questions (see Section \ref{Dataset_Curation}) in which the participant would rank which response addressed the question better. 
We randomly assigned the aspect questions to each prompt instruction, ensuring balanced coverage across all categories, with each question drawn from a different category.
Further, for each instruction and question, we asked the user to provide a brief explanation of their ranking as to why the selected response was chosen over the other. We randomized the order of instructions in the survey for each participant, and in addition we randomized the placement of which candidate response was placed first. The survey was presented using the Qualtrics platform and distributed via email. 

\subsection{LLM Evaluation Experiment Procedure}
\label{sec:llmjudgeframework}

To enable the LLM judge model to rank outputs, we use the widely adopted AlpacaEval framework\footnote{\url{https://tatsu-lab.github.io/alpaca_eval/}}, which identifies the candidate response preferred by the LLM. 
This framework scores a given model based on its \textit{win rate} relative to a baseline model, which is often a well-performing model that is widely accessible and commonly used in practice~\cite{dubois2024alpacafarm, chang2024survey, wang2023aligning}.
AlpacaEval was chosen over other frameworks as it handles some of the common challenges that come with using LLM judges, such as randomizing the order of the two models outputs so there is no positional bias in the evaluation (as was done in our survey with SME participants).

Using the framework, the LLM judge model was provided the same twenty-five prompt instructions, the set candidate responses, and three ranking questions (one general preference question and two aspect questions) as provided to SMEs in the survey. The LLM judge model will rank which model was preferred (response A or response B) for each question. It is important to note that we replace the original AlpacaEval instruction-following dataset with our own curated dataset. 
We also modified the framework to allow for preferences to be evaluated relative to specific aspect questions and instructed the LLM judge model to provide an explanation for its choice between candidate responses.
This is done by simply adding, ``Why is \{preference\} chosen over the other output? Provide a short 2-3 sentence explanation.'' as an additional user message after the initial preference is given.
Additionally, to experiment with the use of expert personas, we modify the AlpacaEval instructions in the system prompt to act as an expert dietitian or clinical psychologist.
An example of our modified AlpacaEval prompt can be found in Table \ref{tab:alpacaeval-template}.

\subsection{Analysis Methods}
\label{sec:analysis}

\subsubsection{SME v. LLM Agreement Rates} We record LLM judge evaluations and SME evaluations on the same set of data.
For each domain, we numerically measure the agreement between LLM judges and SMEs on the same prompt instructions and calculate the percentage of instances where both the LLM judge and SMEs agree.
We also measure the agreement among the SMEs in each domain.

\subsubsection{Qualitative Analysis of Ranking Explanations} To address RQ2, we analyzed the qualitative data collected from both SMEs and LLMs on their explanations for choosing one output over the other. The participants and the LLM provided explanations not only for their overall choice but also for each individual aspect question. We followed a reflexive thematic analysis approach and utilized standard open-coding procedures to uncover underlying themes \cite{braun2019reflecting}. The explanations were reviewed for familiarization with the data, then initial codes were created to categorize the data into themes. During this process, multiple codes could be assigned to a single response to capture the complexity of the feedback. Following discussion of the initial coding, the data was re-coded with the new codes. We provide illustrative examples from both SMEs and LLMs to support and contextualize the final codes. 
We report on the emergent themes and highlight key differences in the explanations provided by the SMEs and the LLMs.

\section{Results}
\label{section:results}

\subsection{SME vs. LLM-as-a-Judge Overall Preferences}
\label{subsec:sme-v-llm-preferences}

When examining the agreement between SMEs and the LLM on \textit{General Preference} questions, we found that SMEs show a relatively low level of agreement with our LLM judge, with $60\%$ agreement in the mental health domain and $64\%$ in the dietetics domain (see Table \ref{tab:llm-vs-sme}). 
By comparison, SMEs agreed with each other $72\%$ in the mental health domain and $75\%$ in the dietetics domain, with an overall agreement of $73\%$ across both domains, establishing a higher baseline for expert agreement.
Further, when adopting the expert persona method in prompt instructions, we observed a $4\%$ improvement in the agreement between the LLM judge and SMEs for both domains in general preference questions, suggesting that personas can enhance agreement between LLMs and SMEs, although agreement levels remain low.
\input{tables/llm-vs-sme-table}

\subsection{SME vs. LLM-as-a-Judge Agreement on Aspect Questions}
\label{subsec:sme-v-llm-aspect-preferences}

Our data shows variation in agreement across the aspect questions and domains, both with and without the expert LLM persona, revealing some notable differences (see Table \ref{tab:llm-vs-sme}). Overall, there was a lower agreement for aspect questions in the dietetics domain with only slight improvement using the expert persona. The agreement in aspect questions tended to be better for mental health SMEs compared to dietetics SMEs with a slight improvement using the expert persona, with the exception of a notable drop in \textit{Clarity} agreement. 

Another noticeable difference is in \textit{Accuracy} agreement, where the LLM shows greater alignment with the mental health SMEs compared to dietetics SMEs. Similar trends of higher agreement are observed in the \textit{Education Context} and \textit{Personalization} categories for the mental health domain compared to the dietetics domain; however, the general model had higher agreement than the expert persona in the dietetics domain. There was also a large difference in \textit{Clarity} agreement, with a much lower alignment in the mental health domain using an expert persona compared to dietetics. 
These results highlight that domain and task complexity affect LLM-SME agreement differently, with the expert persona sometimes improving alignment but not consistently across all aspects.


\subsection{SME vs LLM Explanations: Qualitative Results}
\label{subsec:sme-v-llm-explanations}

In this section, we present the findings on the themes that emerged from the evaluation explanations with illustrative examples from the survey to address RQ2.

\subsubsection{Theme 1: Alignment with Expert Knowledge and Accuracy}



SMEs consistently favored responses that demonstrated \textbf{accurate information}. In the dietetics domain, this included factually correct, up-to-date, evidence-based practices and scientifically supported information that \textbf{aligned with professional standards}. Experts were quick to flag inaccuracies, especially when responses promoted outdated or misleading dietary concepts, misrepresented nutrients, or incorrectly addressed allergens. For example, one dietitian pointed out, \textit{``This statement in response B is not true...carbohydrates alone do not increase the risk necessarily, so the statement is misleading'' (Diet1)}. SMEs also preferred responses that mirrored their clinical recommendations and adhered to industry guidelines. Inaccurate or harmful recommendations were a key concern, with one expert highlighting \textit{``Response B is not a healthy way of thinking. It will NOT benefit someone with diabetes...it is WAY more important to communicate the dangers of the ketogenic diet'' (Diet7)}.

In the mental health domain, accurate responses referenced evidence-based treatments and avoided misinformation, such as promoting non-evidence-based therapies. Responses that reflected the latest research and clinical guidelines were favored, as SMEs recognized the importance of adhering to evidence-based standards. SMEs also appreciated responses that aligned with best practices in therapy and treatment. For example, responses that described therapy approaches and clarified that medications should complement therapy, rather than serve as standalone treatments, were preferred. However, mental health experts expressed concern about responses that included non-evidence-based treatments without proper qualification, or that suggested premature diagnoses based on vague symptoms. Such responses were considered problematic, with one participant noting, \textit{``Response A is bad. Like harmful, bad. Taking a couple nonspecific symptoms and telling someone what they likely have is hugely problematic and reminiscent of the ``Dr. Google'' phenomenon, where patients come in with an inflexible self-perception that may not be accurate'' (Psych6)}.


In contrast to the SMEs' approach, where accuracy and adherence to professional standards were top priorities, the LLM often overlooked some of the harmful or inaccurate aspects of the outputs. Instead, it often repeated specific details provided in the outputs that were focused on following prompt instruction. LLMs missed the critical details identified by SMEs, such as gaps in information or where essential details were missing. 
While SMEs pointed out serious issues with certain responses, the LLM did not flag these inaccuracies, suggesting that it lacked the ability to recognize the potential harm of these recommendations. This discrepancy reveals the primary risk of relying solely on LLMs for the evaluation of accuracy tasks. As LLMs can process and follow instructions, they may overlook critical details and fail to assess the broader implications of the content.

\subsubsection{Theme 2: Clear and Efficient Communication}
SMEs preferred responses with \textbf{clarity and conciseness}, using simple and clear language for a broad audience, and \textbf{terminology} that avoided jargon or overly complex medical terms. SMEs preferred responses that were short, to the point, and avoided unnecessary detail or filler. These types of responses were chosen because they were seen as more digestible for patients with varying levels of health literacy. For example, one explanation from a dietitian states, \textit{``I like response B better, as it is more clear and relays the information in a more organized way. It also states ``milk'' instead of ``dairy'', which is more accurate.'' (Diet4)}. Equally important was the \textbf{structure/flow} of the responses. In both domains, responses that were well-organized, such as breaking down advice in a step-by-step manner or logically grouping related information, were considered to be more effective. For example, in the dietetic domain, the organization of content by prioritizing key factors such as protein and sodium was appreciated for its clarity as were structured lists or menus. Similarly, in the mental health domain, responses deemed to be more clinical or disjointed were less preferred. For example one psychologist notes \textit{``[The response] is short, sweet, and to the point while identifying common symptoms and not overwhelming the reader with too much information (i.e., the reader might have OCD and too much information might cause unnecessary worry)'' (Psych8).}

While the overall communication of the explanations was a significant factor for how the SMEs ranked responses, the LLM also emphasized clarity in its explanations, but in a different manner. The LLM often equated clarity with being more detailed or comprehensive, whereas SMEs prioritized clear, simple language, and the overall flow of the explanation, regardless of how much detail was included. The LLMs version of clarity often did not meet the SMEs standards for how they would structure information to a client, especially when considering the client's lack of understanding.
This difference in what constitutes ``clarity'' could explain the misalignment observed in the Clarity aspect questions, as the SMEs preferred concise and easily digestible information over exhaustive detail. Particularly in the mental health domain, the LLMs missed that an overflow of information, even if inaccurate, may overwhelm or confuse clients and could harm their understanding.


\subsubsection{Theme 3: Referring to Professional Guidance and Tone}
SMEs in both domains favored responses that \textbf{referred to professional guidance} and emphasized the importance of consulting with trained professionals. In the dietitian domain, responses that encouraged people to seek personalized long-term meal plans or advice from registered dietitians or healthcare providers were favored, as these professionals can tailor recommendations to the individual's unique health needs. Similarly, in the mental health domain, responses that directed users to mental health professionals for formal assessments or comprehensive treatment plans were highly regarded, particularly when dealing with complex conditions such as Autism Spectrum Disorder or schizophrenia. For example, \textit{``Both provide information about the approach but the lack of directing to a professional makes it seem like these are activities that the parent should be conducting.'' (Psych1 )}.

For \textbf{tone and framing}, SMEs in both the dietetics and mental health domains valued responses that maintained a positive, empathetic, and encouraging tone while framing information in a way that felt supportive rather than judgmental. In the dietetic domain, responses that provided balanced, non-judgmental nutritional advice such as promoting mindfulness, and conveying ease in adopting healthy habits were preferred. In the mental health domain, tone played a crucial role in how advice was received. Responses that provided empathetic validation, encouraged hope, and normalized difficult emotions were viewed more favorably. The framing of mental health advice in a positive and hopeful manner without making the process seem overly daunting was preferred. For example, a psychologist mentioned that a response \textit{``seemed less pathologizing and had a nice balance of normalizing while also encouraging treatment'' (Psych7)}.

The LLM generally aligned with SMEs, often favoring responses that emphasized the importance of contacting a dietitian or seeking a thorough evaluation from a mental health expert, a recurring theme across both domains. It also occasionally favored responses based on a positive tone, which was in agreement with SMEs' preferences.

\subsubsection{Theme 4: Relevance to Client Needs and Actionable Information}

SMEs preferred responses that had more \textbf{personalization} in both domains particularly when responses considered individual preferences, cultural factors, and specific dietary or mental health needs. In the dietetics domain, recommendations that included culturally relevant foods and addressed user allergies or health conditions were favored. In mental health, preferred responses acknowledged the client's emotional state and tailored coping strategies. 
Mental health SMEs found the inclusion of \textbf{emotional support} important, particularly when individual struggles and offering hope were acknowledged. In addition, responses with \textbf{actionable information} such as concrete steps or implementation strategies, were preferred. In the dietitian domain, SMEs valued specific and detailed advice, such as exact nutrient recommendations or foods, whereas in the mental health domain, step-by-step strategies were appreciated. For example a SME noted \textit{''I think Response A meets the person where they are at -- basically, telling them to go get help and develop a treatment plan.  This is probably what someone in this predicament needs to hear, as opposed to a comprehensive overview of diagnosis/treatment.'' (Psych6).} 

In contrast, the LLM generally favored responses that included actionable or practical steps for the user, but this type of information was sometimes present in both responses, suggesting that LLM's selection may also be influenced by subjective factors, such as the level of detail. Furthermore, while the LLM highlighted personalization by noting when a response was better suited to the type of user mentioned, it often lacked the in-depth rationale provided by SMEs. 


\subsubsection{Theme 5: Depth of Information}
In the dietetic domain, responses that provided a \textbf{comprehensive explanation} or \textbf{detailed information} of the benefits and potential drawbacks of specific diets or foods were appreciated for offering a complete picture. For example, one dietitian stated \textit{``Option B was more educational. It did a better job of teaching why each food was chosen and gave practical tips to help the client learn how to make his own decisions and choose healthy foods to lower his cholesterol.'' (Diet5)}
In the mental health domain, responses that thoroughly explored the emotional, behavioral, and psychological aspects of a condition were preferred. Providing a detailed description of therapeutic interventions offered a more well-rounded response. Those that went beyond simply listing possible treatments and included education about potential diagnoses or addressed both emotional and environmental factors were seen as more holistic. For example a psychologist noted \textit{``A is more thorough and provides a better explanation than B. A talks about the outcomes that come from engaging in CBT and how they can help change the response to anxiety (Psych9)}''.

LLMs often emphasized the importance of comprehensive or detailed responses, as these were the most identified themes in their explanations. While this level of detail sometimes aligned with SME preferences, SMEs were more discerning about how much detail was necessary. They did not always equate more detailed responses with better ones, as they prioritized clarity and relevance over volume of information. This careful balance was something the LLM occasionally missed in its evaluations.


%% file: tables/participant_profiles.tex
\begin{table*}[t!]
\centering
\caption{Profiles of Registered Dietitian (a) and Clinical Psychologist (b) participants. The table includes responses to the following questions: 
1) What is your highest level of education? 
2) How many years of experience do you have in the field? 
3) Have you ever utilized artificial intelligence tools in your practice or research related to psychology, nutrition, or therapy? 
4) How would you rate your familiarity with artificial intelligence and technology in the context of psychology, nutrition, or therapy? 
Familiarity with AI is rated on a scale of 1-5, where 1 = Not Familiar and 5 = Very Familiar.}
\label{tab:combined_profiles}
{\renewcommand{\arraystretch}{1.25}
\resizebox{0.9\linewidth}{!}{%
\begin{tabular}{c|l|l|c|c|c}
\multicolumn{6}{c}{\textbf{(a) Registered Dietitians}} \\ \hline
\textbf{Participant ID} & \textbf{Profession} & \textbf{Education} & \textbf{Years of Experience} & \textbf{Use of AI Tools} & \textbf{Familiarity with AI} \\ \hline
Diet1  & Dietitian  & Bachelor's Degree  & 10 & No  & 3 \\ \hline
Diet2  & Dietitian  & Master's Degree    & 10 & No  & 3 \\ \hline
Diet3  & Dietitian  & Master's Degree    & 18 & Yes & 3 \\ \hline
Diet4  & Dietitian  & Bachelor's Degree  & 3  & No  & 3 \\ \hline
Diet5  & Dietitian  & Master's Degree    & 9  & No  & 2 \\ \hline
Diet6  & Dietitian  & Bachelor's Degree  & 2  & Yes & 4 \\ \hline
Diet7  & Dietitian  & Bachelor's Degree  & 4  & No  & 1 \\ \hline
Diet8  & Dietitian  & Bachelor's Degree  & 4  & No  & 3 \\ \hline
Diet9  & Dietitian  & Master's Degree    & 30 & No  & 2 \\ \hline
Diet10 & Dietitian  & Bachelor's Degree  & 2  & Yes & 4 \\ \hline
\multicolumn{6}{c}{\textbf{(b) Clinical Psychologists}} \\ \hline
\textbf{Participant ID} & \textbf{Profession} & \textbf{Education} & \textbf{Years of Experience} & \textbf{Use of AI Tools} & \textbf{Familiarity with AI} \\ \hline
Psych1 & Psychologist & Doctorate          & 10 & No  & 1 \\ \hline
Psych2 & Psychologist & Doctorate          & 15 & No  & 2 \\ \hline
Psych3 & Psychologist & Doctorate          & 14 & No  & 2 \\ \hline
Psych4 & Psychologist & Professional Doctorate & 26 & No  & 1 \\ \hline
Psych5 & Psychologist & Doctorate          & 6  & No  & 2 \\ \hline
Psych6 & Psychologist & Doctorate          & 10 & No  & 3 \\ \hline
Psych7 & Psychologist & Doctorate          & 11 & No  & 2 \\ \hline
Psych8 & Psychologist & Doctorate          & 7  & No  & 3 \\ \hline
Psych9 & Psychologist & Doctorate          & 38 & No  & 2 \\ \hline
Psych10 & Psychologist & Doctorate         & 15 & No  & 2 \\ \hline
\end{tabular}
}}%
\end{table*}

%% file: tables/llm-vs-sme-table.tex
\begin{table*}[ht]
\centering
\begin{tabular}{|l||c|c||c|c|}
\hline
\multicolumn{1}{|c||}{} & \multicolumn{2}{c||}{\textbf{Dietetics}} & \multicolumn{2}{c|}{\textbf{Mental Health}} \\ \hline
\textbf{Question Type} & \textbf{General Model} & \textbf{Expert Persona} & \textbf{General Model} & \textbf{Expert Persona } \\ \hline
Clarity  & 55\% & 60\% & 70\% & 40\% \\ \hline
Accuracy   & 56\% & 67\% & 80\% & 80\% \\ \hline
Professional Standards  & 80\% & 80\% & 64\% & 73\% \\ \hline
Education Context  & 55\% & 45\% & 60\% & 70\% \\ \hline
Personalization  & 56\% & 44\% & 67\% & 67\% \\ \hline
\textbf{General Preference} & \textbf{64\%} & \textbf{68\%} & \textbf{60\%} & \textbf{64\%} \\ \hline
\end{tabular}
\caption{Comparison of agreement between the General LLM vs. SME and the Expert Persona LLM vs. SME in the Dietetics and Mental Health domains. The \textit{General Preference} question refers to selecting which output is better overall, while the remaining categories focus on specific aspect questions.}
\label{tab:llm-vs-sme}
\end{table*}

%% file: 5_lay_user_followup.tex
\section{Evaluating Lay User Alignment with Large Language Models}
\label{subsec:lay-user-v-llm-preferences}

Our findings in Section \ref{subsec:sme-v-llm-preferences} and \ref{subsec:sme-v-llm-aspect-preferences} reveal a misalignment between the preferences of SMEs and LLMs in evaluating domain-specific tasks. This highlights the necessity for continued SME involvement in the development and fine-tuning of these models.

Current fine-tuning methods for LLMs, such as GPT-4, utilize Reinforcement Learning from Human Feedback (RLHF)~\cite{wang2024comprehensive}. This technique relies on human evaluators to provide feedback, which is then used to align the model's responses with human values.
However, most state-of-the-art LLMs use lay users for RLHF, indicating that the effectiveness is highly dependent on the quality of the feedback provided \cite{christiano2017deep, google2024gemini, openai2024gpt4}.
When SMEs are not involved, often due to the time and cost associated with their expertise, the process may lack the precision required for domain-specific alignment. 
As a result, we hypothesized that LLMs may align more closely with the preferences of lay users, potentially contributing to the misalignment we observed between SMEs and LLMs.

To test this hypothesis and to compare the differences in alignment between SMEs and lay users with the LLM, we conducted a follow-up study in which lay users provided rankings for the same instructions.
We recruited lay users to rank responses to the same survey questions taken by SMEs, answering the prompt \textit{``Which response is better overall?''}. This allowed us to directly compare the agreement between the LLM and both groups, and inform us on potential differences in evaluation preferences between SMEs and lay users. Lay users were recruited through Prolific, a crowdsourcing platform for conducting qualitative studies \cite{prolific}. A total of 10 users were recruited for each domain to complete the pairwise comparison task and were compensated the recommended average rate of \$12.00 per hour. The participant profiles for lay users can be found in Table \ref{tab:lay_profiles} in the Appendix.


We use the same analysis methods to calculate the agreement rates as used in Section \ref{sec:llmjudgeframework} and Section \ref{sec:analysis}. We also compare the agreement rate between the lay user and the general model and between the lay user and the LLM expert persona.

\subsection{Lay User Results}
\begin{figure*}[!ht]
\centering
\includegraphics[width=0.8\textwidth]{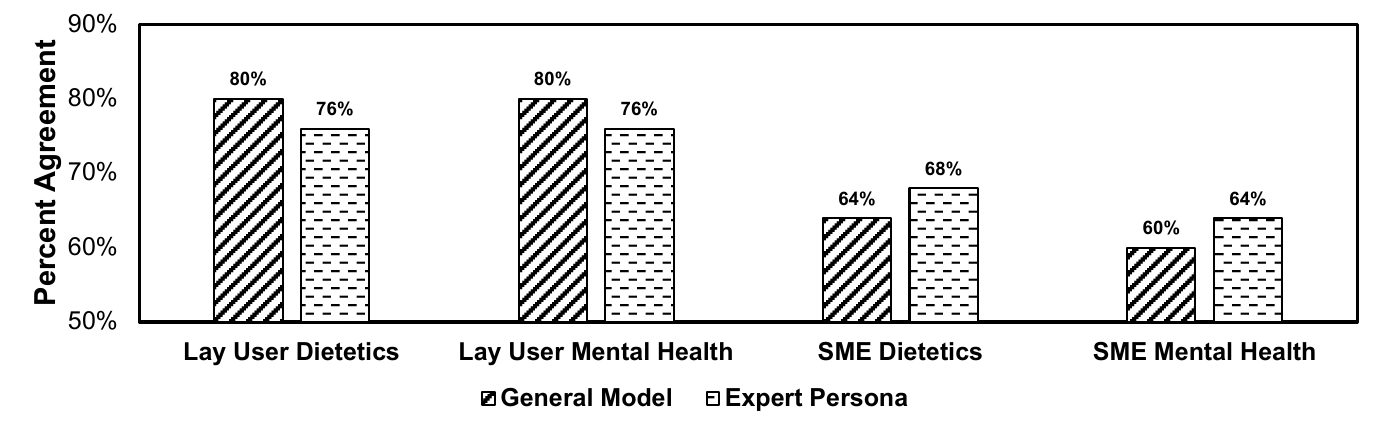}
\caption{Comparison of Agreement between LLM Judge and SMEs versus Lay Users for Dietetics and Mental Health. The lay users show significantly more agreement with the LLM Judge using a standard persona than the LLM Judge using an expert persona. Conversely, the SMEs show more agreement with the LLM Judge using an expert persona than with a standard persona.}
\label{fig:llm-vs-lay-users}
\end{figure*}


The results of our evaluation comparing Lay Users to an LLM judge are shown in Figure \ref{fig:llm-vs-lay-users}.
Interestingly, we observe that the agreement rate between lay users and the LLM is 80\% for both the dietetics and the mental health domain with the default general model. This is a higher agreement than what was observed between the LLM and SMEs, suggesting that the models are more closely aligned with lay user preferences. A two-tailed chi-squared test was conducted, revealing a statistically significant difference (\emph{p} < 0.0001) between the agreement rates of lay users and SMEs with the LLM using the general model and the expert persona LLM in each domain.

In addition, the use of expert personas actually decreases the agreement to 76\% for both dietetics and mental health. This shows the opposite of what we observed in the SME domain, where the use of expert personas improved the agreement rate.
We will discuss the implications of this in Section \ref{subsec:expertpersonadiscussion} in the discussion.

%% file: 6_discussion.tex
\section{Discussion}

This paper contributes new insights into how the LLM-as-a-Judge evaluation approach compares to evaluations conducted by SMEs for domain-specific tasks. Our research highlights the limitations of relying solely on LLMs for evaluation, and the importance of keeping the human, specifically the SME, in the loop for tasks that require expert knowledge. In this section, we explore the potential reasons behind the observed patterns in our findings. We then discuss key design implications that developers and designers should consider when designing and choosing the method for evaluating the quality of LLM outputs for domain-specific tasks.

\subsection{Differences in LLM-as-a-Judge v. SME Evaluation (RQ1 \& RQ2)}

As observed in the results (see Section \ref{section:results}), there is a misalignment between the SMEs and the LLM for both general preference and aspect questions. Additionally, the findings reveal that the alignment between the SMEs and the LLM varies across different domains, even when an expert persona LLM is used. To follow, we highlight the major takeaways in the data and discuss reasoning for the variability.

\subsubsection{Impact of Expert Personas on LLM Alignment with SMEs}
\label{subsec:expertpersonadiscussion}

Instructing the model to adopt an expert persona generally improved alignment with SMEs compared to the general model, indicating that expert-level tasks benefit from personified models for evaluation. 
The importance of the expert persona is further highlighted by the lower agreement between lay users and the expert persona LLM, as lay users may tend to evaluate or rank outputs based on different criteria~\cite{szymanski2024comparing}.
Our findings also show that the decision to use an expert persona or a general-purpose persona should depend on the specific task and domain.


Interestingly, in the dietetics domain, the expert LLM did not improve alignment with SMEs in the \textit{Education Context} and \textit{Personalization} aspect question categories. This suggests that while expert personas may be better suited for tasks requiring strict adherence to professional standards or specialized knowledge, general-purpose models may perform better in tasks that require adaptability and user-centered flexibility. The general-purpose model's broader scope allows it to generate more varied responses, which may resonate better with SMEs in tasks such as education and personalization. 

Another reason for the misalignment may be that expert personas prioritize technical language and specific jargon, which SMEs may not consider to be as effective in educational or practical contexts. This aligns with prior studies where persona agents have become a standard for enabling personalized experiences~\cite{louie2024roleplay}. However, role-playing or personified LLMs often lack datasets that fully represent real human behaviors in specific domains~\cite{tseng2024two, ahn2024timechara}.  
Additionally, many large-scale models are preference-tuned on lay user data, meaning their outputs align more with general preferences rather than those of domain experts. In this case, the model may be heavily influenced by preference tuning/RLHF with limited expert preference data, causing it to lack the specialized focus needed for tasks such as clinical dietetics. 

An interesting finding in the mental health domain is that the expert persona failed to align with SMEs standards in the \textit{Clarity} category. 
From our qualitative data, we determined that SMEs in the mental health domain expressed concerns about overly detailed outputs, which could lead to confusion, self-diagnosis, or even harm. The expert persona may be favoring the more technical outputs with increased vocabulary, but failed to detect the risks and potential misuse that SMEs typically flag.


\subsubsection{Variability in LLM v. SME Alignment across Domains and Aspect Categories}
In both the dietetics and mental health domains, we observed high variability in the alignment between SMEs and the LLM depending on the aspect question. 
For example, in the mental health domain, the SMEs showed higher alignment in questions related to \textit{Accuracy} and \textit{Educational Context}.
This variability suggests that LLMs trained on mental health tasks may benefit from the availability of well-structured resources in the training data, such as research papers and educational materials, which contributes to higher alignment with SMEs. 
In contrast, the dietetics domain may face challenges such as conflicting dietary advice, misinformation, and rapidly evolving guidelines, making it harder for LLMs to achieve close alignment with SMEs in these areas. Popular diet fads and varied nutritional recommendations further complicate this alignment. Knowing this means that we need human experts to pay closer attention to certain aspects, particularly in domains where LLMs are prone to lower alignment, such as dietetics.

The SMEs in the dietetics domain showed lower alignment with \textit{Personalization} questions compared to the mental health domain, suggesting that addressing dietary needs may involve greater complexity. In mental health, personalization often involves broader, less detailed approaches, whereas in dietetics, it requires more specific, actionable nutrition-related recommendations. Our results show that mental health evaluations were also heavily influenced by the level of emotional support provided in responses, while dietary experts focused more on the practicality and specificity of the recommendations offered. 



In addition, the dietetics domain showed slightly higher alignment with SMEs compared to the mental health domain on \textit{Professional Standard} questions. This may be because questions in the mental health domain often focused on determining which responses demonstrated an empathetic or encouraging tone that was more in line with accepted expert standards. The lower misalignment in mental health indicates that the LLM may struggle to consistently apply empathy in a manner that aligns with professional standards, highlighting a potential challenge in effectively balancing technical accuracy with emotional sensitivity.


\subsubsection{SMEs' Unique Contributions in Feedback Compared to LLMs}
When analyzing the qualitative results (see Section \ref{subsec:sme-v-llm-explanations}), we found that SMEs provided much more specific and unique context compared to the LLMs when addressing the explanation behind the evaluation ranking. SMEs' feedback was much more detailed and context-specific, whereas LLMs tended to provide generalized explanations. In many cases, LLMs appeared to repeat information from the instruction or the output itself with limited new information being contributed. 


On the other hand, SMEs were shown to contribute new information and unique insights that they likely drew upon from their knowledge or experience instead of reiterating the information provided. This could be attributed to the pattern-based learning of LLMs inherent in autoregressive language modeling. Furthermore, the limited contextual understanding and lack of domain-specific expertise in LLMs contribute to their tendency to generalize. This overreliance on the instruction, combined with the insufficient use of parametric knowledge, restricts the LLM’s ability to provide the depth of feedback seen in SMEs.



\subsubsection{Lay User Alignment with LLMs in General Preference Tasks}
Interestingly, our findings in Section \ref{subsec:lay-user-v-llm-preferences} also reveal that lay users tend to have higher alignment with LLMs than do SMEs for general preference tasks. This suggests that LLMs may be preference-tuned using data primarily from lay users. As a result, the models may align more with lay preferences than with expert knowledge. This can lead to a divergence in performance when evaluating tasks that require specialized expertise, emphasizing the need to carefully consider the training and evaluation process for domain-specific tasks.






\subsection{Implications for Evaluating Systems in Domain-Specific Tasks}

Based on the patterns identified from our analysis, we present the following design implications to guide researchers and developers in evaluating systems for domain-specific tasks.

\subsubsection{Incorporate SME-in-the-Loop for Personalized and Complex Evaluations}
Researchers and developers should integrate a SME-in-the-loop approach for evaluations that involve complex decision-making. SMEs can provide critical insights, especially when LLMs are deployed in domains that demand specialized expertise. The SME-in-the-loop approach ensures that the system's output is validated and fine-tuned to meet high standards in complex tasks, preventing errors that could arise from overreliance on generalized LLM outputs. Although integrating SMEs can be costly and time-consuming, involving SMEs in pairwise rankings or iterative review processes can help identify subtle errors and misalignments that automated systems might miss. This human-in-the-loop approach is especially valuable in tasks where the precision of recommendations is essential, such as healthcare, where the consequences of mistakes can be significant.

However, LLM judges are well suited for conducting large-scale evaluations across many models \cite{zhu2023judgelm}. By using LLMs to evaluate performance on large datasets and across numerous model variations, the worst-performing models can be eliminated early on. SMEs can then be brought in to evaluate the remaining top-performing models, ensuring expert-level insights are applied efficiently and cost-effectively. This hybrid approach allows the scalability and speed of LLM evaluations while maintaining high standards for complex, high-stakes decisions through the involvement of SMEs.





\subsubsection{Domain-Specific Evaluation Frameworks and Data Curation}

Given the variation in performance across dietetics and mental health domains, a design implication is the need for LLMs to be evaluated within a domain-specific framework. The fact that alignment varied between categories such as \textit{Accuarcy} and \textit{Education Context} suggests that a one-size-fits-all evaluation approach may overlook critical domain-specific weaknesses. When designing evaluations for systems that operate across multiple domains, each domain should have a tailored evaluation framework, similar to the one used in our study. 

Domain-specific evaluation frameworks can be used to assess the alignment between SMEs and LLMs in certain task criteria.
Researchers or designers can curate a domain-specific dataset that addresses the unique characteristics and criteria of each field. By utilizing a common pairwise comparison method \cite{li2023alpacaeval} between SMEs and LLMs, we can pinpoint critical evaluation criteria, allowing SMEs to focus on areas where they are most needed, thereby saving time and reducing costs.
By testing the model against these curated datasets, developers can better understand the strengths and weaknesses of LLMs in each domain and refine the models to improve their alignment with expert standards.

\subsubsection{Tailor LLM Personas to Domain-Specific Expertise for Improved Alignment}
 From other literature, we see that personas are commonly used to generate more personalization and expert dialogue~\cite{liu2024personaflow, salemi2023lamp}. Our study expands on these findings by suggesting that LLM personas should be carefully tailored to the domain-specific expertise required for the task. For domains such as dietetics or mental health, where expert-level precision is critical, training LLMs to adopt specialized personas based on domain-specific guidelines and knowledge is essential. These expert personas enable the model to offer more accurate, professional, and context-aware outputs, particularly for high-stakes or regulatory-driven domains. However, for tasks that require adaptability and general user interaction, general-purpose models may be more effective. This suggests that more domain-specific training, either using datasets such as clinical guidelines or therapeutic frameworks or through interactive human-AI alignment efforts~\cite{gebreegziabher2024supporting, gebreegziabher2024leveraging}, may be necessary to improve alignment in categories where expert knowledge is vital. Developers should leverage domain-specific datasets and expert personas to fine-tune LLMs for optimal performance in judging specialized tasks.




%% file: 7-conclusion.tex
\section{Future Work \& Limitations}
 Our work has some limitations that open the door to exciting future work. First, we do not thoroughly explore the correlation between pre-training data content and downstream performance on domain-specific tasks. Future work has the opportunity to examine how differences in pre-training data influence judgment alignment in various domains, especially where misinformation or conflicting guidelines are prevalent, as in dietetics.

Additionally, the models used in this study were not further tuned on the pairwise comparisons or the qualitative explanations provided by SMEs. This is a common practice in LLM research, but involves significant costs and resource demands in fine-tuning on specialized feedback.
Future work has the opportunity to explore further tuning LLMs using this feedback, allowing the models to learn from SME judgments and refine their outputs. We believe that this approach could help significantly reduce the observed misalignment in specific domains.

Lastly, our study is limited to only the two specific domains of dietetics and mental health due to the high cost associated with collecting annotations from SMEs.
Although these fields are representative of the kinds of complex tasks that require expert reasoning, we invite future work to expand this evaluation to other fields that require domain expertise (e.g., qualitative analysis~\cite{gebreegziabher2023patat, gao2024collabcoder}, creativity~\cite{suh2024luminate, zhang2023visar}, UX~\cite{lu2022bridging, lu2024ai, lu2024flowy}, and academic research~\cite{kapania2024m}). Future work may also explore how LLMs perform when evaluating other domain-specific tasks to explore the benefits and weaknesses of evaluations that are unique to other fields.  

\section{Conclusions}
This paper explores the differences between SMEs and LLM-as-a-judge in evaluating domain-specific tasks that require expert knowledge. Our findings reveal that it is important to design evaluation systems to keep SME-in-the-loop, and our results highlight the strengths and weaknesses identified in the evaluations between LLM and SMEs. We also provide an empirical understanding of the factors that SMEs care about during the evaluation process that differ from LLMs. 
Our analysis suggests implications that should be considered for evaluating systems in domain-specific tasks, such as integrating SME-in-the-loop and preference tuning or RLHF, developing frameworks on where to best integrate expertise, and using LLM expert personas to improve judgement alignment. 
Ultimately, this work contributes to the discussion on SME-driven evaluation, offering insights to enhance alignment between human experts and LLMs in interactive systems.

%% file: 8-acknowledge.tex
\section*{Acknowledgements}
This work was supported by the Agriculture and Food Research Initiative grant no. 2021-67022-33447/project accession no. 1024822 from the USDA National Institute of Food and Agriculture and by the Office of Naval Research N00014-22-1-2507.

%% file: 9-appendix.tex
\appendix
\section{Appendix}

\begin{table}[htbp]
    \small
    \centering
    \renewcommand{\arraystretch}{1.4}
    \begin{tabular}{p{14cm}}
    \lstset{
        basicstyle=\ttfamily\small,
        breaklines=true,
        numbers=none,
        frame=single,  
        framerule=0.8pt,  
        framesep=5pt,  
        backgroundcolor=\color{white},  
        rulecolor=\color{black}  
    }
    \begin{lstlisting} 
<|im_start|>system
You are a highly efficient assistant, who evaluates and selects the best large language model (LLMs) based on the quality of their responses to a given instruction with respect to a particular aspect question. You are also a clinical psychologist. This process will be used to create a leaderboard reflecting the most accurate and human-preferred answers.
<|im_end|>
<|im_start|>user
I require a leaderboard for various large language models. I'll provide you with prompts given to these models and their corresponding outputs. Your task is to assess these responses, and select the model that produces the best output from a human perspective with respect to a particular aspect question.

## Instruction

{
    "instruction": """{instruction}""",
    "aspect_question": """{aspect_question}"""
}

## Model Outputs

Here are the unordered outputs from the models. Each output is associated with a specific model, identified by a unique model identifier.

{
    {
        "model_identifier": "m",
        "output": """{output_1}"""
    },
    {
        "model_identifier": "M",
        "output": """{output_2}"""
    }
}

## Task

Evaluate the models based on the quality and relevance of their outputs, and select the model that generated the best output with respect to the aspect question. Answer by providing the model identifier of the best model. We will use your output as the name of the best model, so make sure your output only contains one of the following model identifiers and nothing else (no quotes, no spaces, no new lines, ...): m or M.

## Best Model Identifier
<|im_end|>

    \end{lstlisting}
    \end{tabular}
    \caption{Prompt template from AlpacaEval adapted with personification and to address individual aspect questions.}
    \label{tab:alpacaeval-template}
\end{table}

\input{tables/lay_user_participant_profile}

%% file: tables/lay_user_participant_profile.tex
\begin{table*}[t!]
\centering
\caption{Profiles of Lay User Diet (a) and Lay User Psych (b) participants. The table includes responses to the following information: 
1) Age 
2) Sex 
3) Ethnicity 
4) Education 
5) Whether they are a client of related services.}
\label{tab:lay_profiles}
{\renewcommand{\arraystretch}{1.25}
\begin{tabular}{c|c|c|p{2.5cm}|l|c} 
\multicolumn{6}{c}{\textbf{(a) Lay User -- Dietetics}} \\ \hline
\textbf{Participant ID} & \textbf{Age} & \textbf{Sex} & \textbf{Ethnicity} & \textbf{Education} & \textbf{Client?} \\ \hline
Lay User\_Diet 1  & 20  & Female  & White        & Some College        & No  \\ \hline
Lay User\_Diet 2  & 35  & Female  & Other        & Bachelor’s Degree   & Yes \\ \hline
Lay User\_Diet 3  & 32  & Male    & White        & Bachelor’s Degree   & No  \\ \hline
Lay User\_Diet 4  & 30  & Male    & Other        & Bachelor’s Degree   & No  \\ \hline
Lay User\_Diet 5  & 28  & Female  & White        & Master’s Degree     & No  \\ \hline
Lay User\_Diet 6  & 44  & Male    & Mixed        & High School         & No  \\ \hline
Lay User\_Diet 7  & 40  & Male    & Black        & Master’s Degree     & Yes \\ \hline
Lay User\_Diet 8  & 51  & Female  & Asian        & High School         & No  \\ \hline
Lay User\_Diet 9  & 64  & Male    & White        & High School         & No  \\ \hline
Lay User\_Diet 10 & 24  & Female  & Black        & Bachelor’s Degree   & No  \\ \hline
\multicolumn{6}{c}{\textbf{(b) Lay User -- Mental Health}} \\ \hline
\textbf{Participant ID} & \textbf{Age} & \textbf{Sex} & \textbf{Ethnicity} & \textbf{Education} & \textbf{Client?} \\ \hline
LayUser\_Psych1  & 25  & Female  & Asian        & Bachelor’s Degree   & Yes \\ \hline
LayUser\_Psych2  & 63  & Female  & White        & Doctorate           & No  \\ \hline
LayUser\_Psych3  & 58  & Female  & White        & Bachelor’s Degree   & Yes \\ \hline
LayUser\_Psych4  & 26  & Male    & White        & Some College        & No  \\ \hline
LayUser\_Psych5  & 37  & Male    & White        & Bachelor’s Degree   & Yes \\ \hline
LayUser\_Psych6  & 22  & Male    & White        & Bachelor’s Degree   & No  \\ \hline
LayUser\_Psych7  & 33  & Female  & Mixed        & Master’s Degree     & Yes \\ \hline
LayUser\_Psych8  & 52  & Male    & White        & Bachelor’s Degree   & No  \\ \hline
LayUser\_Psych9  & 21  & Male    & White        & High School         & No  \\ \hline
LayUser\_Psych10 & 30  & Male    & Black        & Bachelor’s Degree   & No  \\ \hline
\end{tabular}
}
\end{table*}